\documentclass[journal,comsoc]{IEEEtran}
\usepackage[T1]{fontenc}% optional T1 font encoding

\usepackage{cite}
\ifCLASSINFOpdf
   \usepackage[pdftex]{graphicx}
   \graphicspath{{img/}}
\else
\fi
\usepackage{amsmath}
\usepackage[cmintegrals]{newtxmath}
\usepackage{bm}

\usepackage[numbers,sort&compress]{natbib}

\usepackage{enumitem}

\begin{document}
\title{A Computation Offloading Incentive Mechanism with Delay and Cost Constraints under 5G Satellite-ground IoV Architecture}

\author{Minghui LiWang, Shijie Dai, Zhibin Gao, Xiaojiang Du, Mohsen Guizani, Huaiyu Dai% <-this % stops a space
\thanks{\hspace{-1em}Minghui LiWang, Shijie Dai and Zhibin Gao (corresponding author) are with Xiamen University.}% <-this % stops a space
\thanks{\hspace{-1em}Xiaojiang Du is with Temple University.}% <-this % stops a space
\thanks{\hspace{-1em}Mohsen Guizani is with University of Idaho.}
\thanks{\hspace{-1em}Huaiyu Dai is with North Carolina State University.}}


\markboth{Journal of \LaTeX\ Class Files, vol. xx, no. xx, March~2018}%
{Minghui LiWang \MakeLowercase{\textit{et al.}}: An Incentive Computation Offloading Mechanism with Delay and Cost Constraints under 5G Satellite-ground IoV Architecture}



\maketitle

\begin{abstract}
The 5G Internet of Vehicles has become a new paradigm alongside the growing 
popularity and variety of computation-intensive applications with high 
requirements for computational resources and analysis capabilities. 
Existing network architectures and resource management mechanisms may not sufficiently 
guarantee satisfactory Quality of Experience and network efficiency, mainly suffering from 
coverage limitation of Road Side Units, insufficient resources, and unsatisfactory computational 
capabilities of onboard equipment, frequently changing network topology, 
and ineffective resource management schemes. To meet the demands of such 
applications, in this article, we first propose a novel architecture by 
integrating the satellite network with 5G cloud-enabled Internet of Vehicles to 
efficiently support seamless coverage and global resource 
management. A incentive mechanism based joint optimization problem of 
opportunistic computation offloading under delay and cost 
constraints is established under the aforementioned framework, in which a 
vehicular user can either significantly reduce the application completion time by 
offloading workloads to several nearby vehicles 
through opportunistic vehicle-to-vehicle channels while 
effectively controlling the cost or protect its own profit by providing compensated computing service. As the optimization problem is non-convex and NP-hard, simulated annealing based on the Markov Chain Monte Carlo as well as the metropolis algorithm is applied to solve the 
optimization problem, which can efficaciously obtain both high-quality and 
cost-effective approximations of global optimal solutions. The 
effectiveness of the proposed mechanism is corroborated 
through simulation results.
\end{abstract}

\begin{IEEEkeywords}
Satellite-ground networks, Internet of Vehicles, Computation offloading, 5G networks
\end{IEEEkeywords}

% \fi
\IEEEpeerreviewmaketitle

\section*{Introduction}
%\IEEEPARstart{T}{his} demo file is intended to serve as a ``starter file''
%for IEEE Communications Society journal papers produced under \LaTeX\ using
%IEEEtran.cls version 1.8b and later.
%% You must have at least 2 lines in the paragraph with the drop letter
%% (should never be an issue)
%I wish you the best of success.

\IEEEPARstart{I}{n} past decades, the role that satellite communications technologies can 
play in the forthcoming 5G Internet of Things (IoT) has been revisited. Several feasible 
proprietary solutions and recent advances in satellite networks such as High 
and Ultra High Throughput Systems (UHTS), which are built on Extremely High 
Frequency (EHF) bands and free space optical links, have ushered in a new era 
where the satellite can be expected to play a fundamental role in facilitating 
more demanding broadcast/broadband services, effective resource and  
mobility management, and achieving a large population of on-ground mobile users 
such as cellphones, tablets, and smart cars~\cite{1,2}. With the development of wireless technologies and applications \cite{3,4,5,6,7}, interconnected smart cars are considered as the next 
frontier in automotive revolution, whereas driverless vehicles have become a 
future trend with the number of connected vehicles predicted to reach 250 
million by 2020~\cite{8}. Moreover, many technological advancements such 
as on-board cameras and embedded sensors open up new application types with 
advanced, computation-intensive features such as personalized automatic navigation, 
accident alerts, and 3D map modeling.

Nowadays, the Internet of Vehicles (IoV) formed mainly by connected vehicles, 
roadside infrastructures, as well as pedestrians has faced with many 
challenges principally owing to high vehicular mobility, low transmission 
rates, local resource limitations and the computational capabilities of onboard 
equipment, leaving vehicles struggling to complete computation-intensive 
applications locally while ensuring a satisfactory Quality of Experience (QoE). To enhance 
users' QoE despite increasing demands on applications, a cloud-enabled 
framework is introduced that allows computation-intensive applications 
to be executed either partially or fully on a cloud computing server such 
as a location-fixed cloud computing center and nearby Mobile Device Computing (MDC) servers (e.g., 
neighboring vehicles as vehicular clouds). These developments efficiently alleviate resource 
constraints and ease the heavy execution burden of vehicles by migrating 
part of the workload to resource-rich surrogates. Two commonly used platforms 
in IoV are Dedicated Short-Range Communications (DSRC) based 802.11p 
networks and LTE cellular networks; however, both have difficulty supporting 
high mobility, and frequent handovers associated with 
different Road Side Units (RSUs) and Base Stations (BSs) become problematic as networks 
grow denser. Fortunately, with the booming technology revolution accompanying the 
advent of 5G, high data-rate transmission capabilities along with
soft handover as well as reduced latency and high reliability can be 
provided to strongly support cloud-enabled IoV. These developments are especially advantageous for 
edge cloud computing and different communication modes including
vehicle-to-vehicle (V2V) and vehicle-to-infrastructure (V2I), at a high data-rate level.

In spite of the effectual part played by 5G cloud-enabled IoV, the 
demands for different vehicular applications have exploded given the prospect
that the data transmission, storage, and processing capacity of information systems 
are expected to grow by 1000 times over the next 10 years. Similarly, 
information system performance is anticipated to be over 1000 
times higher for the sake of achieving green development. Furthermore, 
the offloading mechanisms of computation-intensive applications are  
vulnerable to signal coverage and viewpoint limitations in addition to being 
highly susceptible to ineffective resource management. As a result, 
the increasingly complex network environment in IoV requires more powerful 
centralized management to overcome the coverage limit, integrate 
multiple resources, and improve global network effectiveness to the greatest 
possible. 

An efficient and future-proof complementary solution to 5G terrestrial IoV 
is the 5G satellite-ground cooperative system, which contains open 
architecture based on Software Defined Networking (SDN) and Network 
Function Virtualization (NFV) technologies~\cite{9,10}. To fulfill diverse 
requirements, the role of the satellite-ground cooperative system can be 
fundamental to reaching areas where terrestrial IoV services are 
limited as well as managing and optimizing the global system performance by taking a 
macroscopic view. By combining a satellite~network with a 5G ground 
IoV system and adopting the efficient fusion of computing, 
communication, and control technology, the satellite-ground cooperative system 
can provide seamless signal coverage and better support for real-time 
perception, dynamic control, and information services to manage large numbers
 of vehicular applications. For~instance, the Intelsat satellite 
antenna (i.e., mTenna) can be embedded into the roof of a vehicle to acquire 
satellite signals even without RSU coverage. Toyota's Mirai Research 
Vehicle equipped with mTenna can provide on-the-move services, which has been
demonstrated to achieve a data rate of 50 Mb/s~\cite{11}. In regions with 
signal coverage, vehicular information such as location, velocity, accident 
warnings as well as application execution requests and resource shortage 
warnings can be gathered by RSUs or BSs and then transmitted to the central controller acted by satellites
 via satellite-ground stations, which can 
macroscopically rationalize both resource deployment and mobility 
management. This arrangement provides appropriate resource allocation guidance for every 
vehicular user to ultimately improve the efficiency of the entire 
network.

In this article, which targets computation-intensive 
applications with high QoE while protecting benefits for all vehicular 
users, we establish an integrated satellite-ground cooperative IoV 
architecture by considering  advanced~5G technologies such as V2V 
communication, cloud computing, and SDN, under which we propose an 
incentive mechanism based joint optimization framework of opportunistic 
computation offloading under delay and monetary cost constraints. The satellite 
network is regarded macroscopically as an integrated control center where 
most functions, control, and management capacities are supported 
through SDN-based interfaces. Computation resources are virtualized into pools with 
resource blocks (RBs) that can be mapped to physical resources. Vehicular 
users are classified into two categories: the buyers, who have
computation-intensive applications waiting to be executed under limited 
resources and computational capability, and the sellers who have idle 
computational resources that can be lent to the buyers for profits. The main duty of the satellite network is to identify a 
mechanism that can rationally allocate resources of the sellers while 
guaranteeing their profits so as to motivate them to provide more service in the 
future from a global perspective. Under the guidance of a central controller, 
one buyer can appropriately assign its workload to multi-sellers 
through several opportunistic one-hop V2V channels, significantly reducing 
application duration and controling costs while improving wireless spectrum 
utilization. Sellers are able to derive appreciable income while increasing the 
utilization of idle resources. The separation between the control signaling and the data 
plane effectively realizes the flexible management of all network 
traffic.

Based on the constraints of delay, cost, and opportunistic contact, we 
establish a novel mathematical model in which the simulated annealing algorithm is 
utilized to identify near-optimal solutions for the offloading 
data rate and the knockdown price per RB. The practical effectiveness and 
efficiency of the proposed mechanism are corroborated through simulations 
and experiments. 

The rest of this article is organized as follows: after describing  
related work in both computation offloading and integrated satellite-ground 
networks, the architecture of the integrated 5G satellite-ground 
cloud-enabled IoV is introduced. The problem definition and system models of the 
proposed mechanism are presented in the following section. Then, the joint 
optimization framework of opportunistic offloading under delay and 
cost constraints is presented in detail, after which a solution based on 
simulated annealing is designed. Finally, we analyze the performance of the 
mechanism and present numerical results before concluding the article.

\section*{Related work}

The integration of satellite and 5G IoV has attracted considerable scholarly attention in recent years as a novel and judicious idea for supporting diverse services in a seamless, efficient, and global manner. 
Several studies have focused on different issues including framework design and coverage extension,  
but comparatively few are attentive in the issue of global resource management.
Computational resources warrant particular attention for reasons that both computational and analytical  
capabilities have come to occupy increasingly important positions in different applications due to 
swiftly developing advanced technologies such as 3D modeling, artificial 
intelligence (AI), and augmented~reality~(AR). In this section, we conduct 
a comprehensive investigation on both 
computation offloading and integrated satellite-ground 
architecture in IoV from the perspectives of motivation and feasibility.

\subsection*{Motivations and Feasibility of Computation Offloading in IoV}

With the rapid development and widespread popularity of IoV, computation-resource-hungry applications have become increasingly necessary among 
vehicular users, especially in terms of historical traffic analysis, personalized 
navigation, dangerous driving warnings, etc. All the above mentioned 
applications require significant computational resources 
and strong computing abilities in order to support big data analysis. 
However, the resource limitations and unsatisfactory computational capabilities of 
onboard equipment result in undesirable Quality of Service (QoS) and QoE. 
An innovative solution to address this issue is cloud-enabled IoV, established over the past few years to allow  
computation-intensive applications to be executed either in part or in full on 
reliable cloud computing servers. The first widely used paradigm was the remote 
cloud~\cite{12,13}, but it resulted in huge transmission delays, serious signal degradation, and low 
reliability~\cite{14} due to variability in the network topology, wireless network capacity 
limitations, and delay fluctuations in transmission on the backhaul and backbone networks. Then, Mobile Edge Computing (MEC) technology at the edge of pervasive Radio Access Networks 
(RAN) in close proximity to vehicles became popular~\cite{20} but continued to be plagued by 
resource constraints as well as RSU radio coverage limits. To 
overcome communication range constraints and make full use of the 
opportunistic contacts between moving vehicles, vehicular clouds open up new 
schemes that allow resource exchange between vehicles through one-hop V2V 
channels by capitalizing on economic efficiency. In a vehicular cloud 
scenario, a vehicle can flexibly play the part of either a seller providing 
computing services while charging a certain fee or a buyer who has a 
computation request to be executed. Literature~\cite{15} investigated a 
cloud-assisted vehicular network architecture in which each cloud had its own 
features, and a corresponding optimal scheme was obtained by solving a 
Semi-Markov Decision Process aimed at maximizing the system's expected 
average reward. To improve network capacity and system computing capabilities, 
authors in~\cite{16} extended the original Cloud Radio Access Network (C-RAN) to 
integrate local cloud services and provide a low-cost, scalable, 
self-organizing, and effective solution called enhanced C-RAN with essential 
technologies of device-to-device (D2D) and heterogeneous networks based on a matrix game 
theoretical approach. Although several studies have solved the problem 
to some extent, limitations can still remain in the strict requirements regarding 
contact and inter-contact duration between vehicles and effective resource 
management solutions from macroscopic and longer-term~perspectives for 
sustainable~development (e.g., incentive~mechanisms to protect sellers). Therefore, attempts to logically improve both resource management and the system framework while insuring the rights of all users are foreseen to be urgent. 

\subsection*{Motivations and Feasibility of Integrated Satellite-ground 5G IoV}

Generally, a satellite network is composed of several satellites, ground stations (GSs), and network operations control centers (NOCCs), and usually provides services for navigation, emergency rescue, communication/relaying, and global resource and geographical information management. On the basis of altitude, 
satellites can be categorized into either geostationary orbit (GSO), medium Earth 
orbit (MEO), or low Earth orbit (LEO) satellites~\cite{11}. Owing to many advantages of satellite networks such as wide-area coverage, reliable access providing mechanism, global information coordination and broadcasting/multicasting capability on supporting massive users, the motivations for integrating satellite networks with IoV can be summarized as follows: 

\begin{itemize}
	\item To provide a kind of seamless coverage service to overcome the coverage and distribution limitations of RSUs in sparsely populated rural areas (e.g., mountainous areas and the desert).
	\item Different applications in IoV cannot be served efficiently by a single technology; hence, the convergence of various networks is likely to become a major trend in the future.
	\item Potential network congestion may occur even in suburban areas with high spatial-temporal dynamics in traffic loads due to vehicular mobility.
	\item A lack of macroscopic management to improve overall network efficiency and resource utilization without satellite networks.
\end{itemize}

Several proprietary solutions and open standards have been developed to 
enable data broadcasting via satellite to mobile users over the years. A 
plan called "Free Space Optical Experimental Network Experiment 
Program (FOENEX)" was implemented by the 
Defense~Advanced~Research~Projects Agency (DARPA), which brought the data 
transmission rate of Mobile Backbone Communication Networks (MBCNs) to 
10Gb/s and even 100Gb/s when using Wavelength Division Multiplex (WDM) 
technology between 2010 and 2012. For satellite networks, Free Space Optical 
Communication (FSO) is the primary choice of MBCNs but fails to cut 
through clouds. Attempting to design and execute an airborne communication 
link with equivalent capacity and optical communication distance, 
DARPA prepared a plan entitled "100G RF Backbone Program" in early 2013~\cite{17}. 
Moreover, some existing researchful studies have focused on different issues based on 
the satellite-ground vehicular network framework. Authors in~\cite{10} examined a 
use case for the realization of end-to-end traffic engineering in a combined 
terrestrial-satellite network used for mobile backhauling. Literature~\cite{18} 
proposed an analytical assessment of the cooperation limits in the 
presence of both a satellite and a terrestrial repeater (gap filler) and 
derived exact expressions and closed-form lower bounds on coverage in a 
setup of practical interest by using the max-flow min-cut theorem. 
Furthermore, they studied a practical implementation of the Random Linear 
Network Coding (RLNC) cooperative approach for the Digital Video 
Broadcasting-Satellite services to Handheld (DVB-SH) standard. All 
previous research have proved that advanced features from different segments can be exploited to support multifarious vehicular applications and scenarios in an efficient manner through interworking, which spurred opportunities to integrate satellite networks with IoV successfully.

\begin{figure*}[!t]
	\centering
	\includegraphics[width=0.9\linewidth]{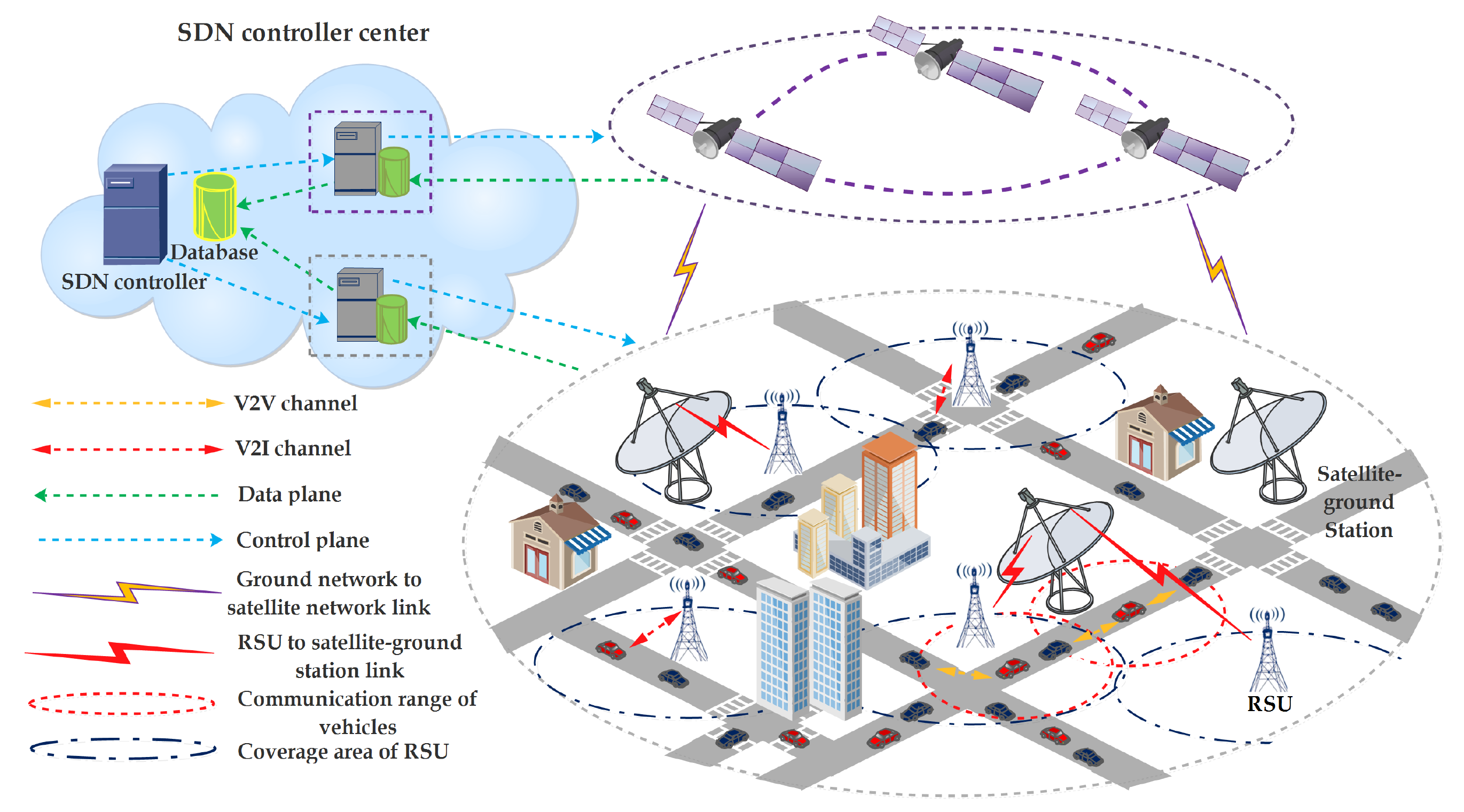}
	\caption{Integrated architecture of the satellite network and 5G cloud-enabled IoV.}
	\label{fig01}
\end{figure*}

\section*{Integrated architecture for the satellite network and 5g cloud-enabled IoV}

SDN is regarded as an emerging network architecture in 5G that separates 
the control plane from the data plane, introduces logically centralized control 
with a global and macroscopic view of the network, and facilitates network 
programmability/reconfiguration through open interfaces~\cite{11}. In 
this section, we provide an introduction to the proposed 
satellite-ground 5G IoV. As shown in Fig. 1, the 5G satellite-ground IoV 
mainly incorporates a satellite network segment and the terrestrial IoV segment. 
SDN controllers exist either on powerful servers or cloud computing centers 
of both the satellite network and terrestrial networks and can have different 
functions given the diverse characteristics of each segment. We develop 
hierarchical SDN controllers to coordinate various features and 
operations of different segments while generating gradational information  
from microscopic to macroscopic perspectives. Cases such as vehicular velocity and direction information can be collected by RSUs.
 Local traffic information (e.g., traffic density and accident warning) can be transmitted from RSUs to the satellite-ground stations, 
and macroscopic perceptions will be gathered in the satellite network 
to facilitate decision making at different tiers of the SDN 
controllers of each segment. In remote regions like mountainous areas and 
deserts without terrestrial signal coverage, vehicles can get contact with 
the satellite network either directly or through the satellite-ground station by installing 
a specific transceiver antenna so as to enjoy seamless coverage service.

 It is noteworthy that vehicles will not encounter interference from the various levels of service 
based on the premise that network slicing technology is performed in each 
segment to partition resources of the entire network  into various slices for 
different services, wherein operations are executed in isolation so as not to interfere with each other.

\section*{Problem definition and system models}

In this section, we provide a detailed overview of the problem definition and 
system models. For one snapshot at time $t$, the whole terrestrial IoV can 
be divided into several cooperative groups, each containing one buyer and 
several sellers who can interact with the buyer through an opportunistic 
one-hop V2V channel. A buyer can offload partial application data to sellers 
and obtain resultant feedback after the sub-applications are completed. Notably, one seller can provide service for different buyers 
simultaneously, and these services are independent without affecting each 
other. The main duties of a central controller for each group can be defined 
as follows. Firstly, a scheduling of a computation-intensive application which can be explained as the 
appropriate data size allocated to one seller in order to reduce the 
application duration under cost constraints; then, the unit knockdown price 
of the computational RB for each seller is decided while controlling 
the buyer's monetary cost and guaranteeing the seller's profits as well as 
willingness to provide idle resources in the future. The schematic diagram of 
computation offloading in individual groups and interactions in local areas 
is shown in Fig. 2. 

\begin{figure*}[!t]
	\centering
	\includegraphics[width=0.9\linewidth]{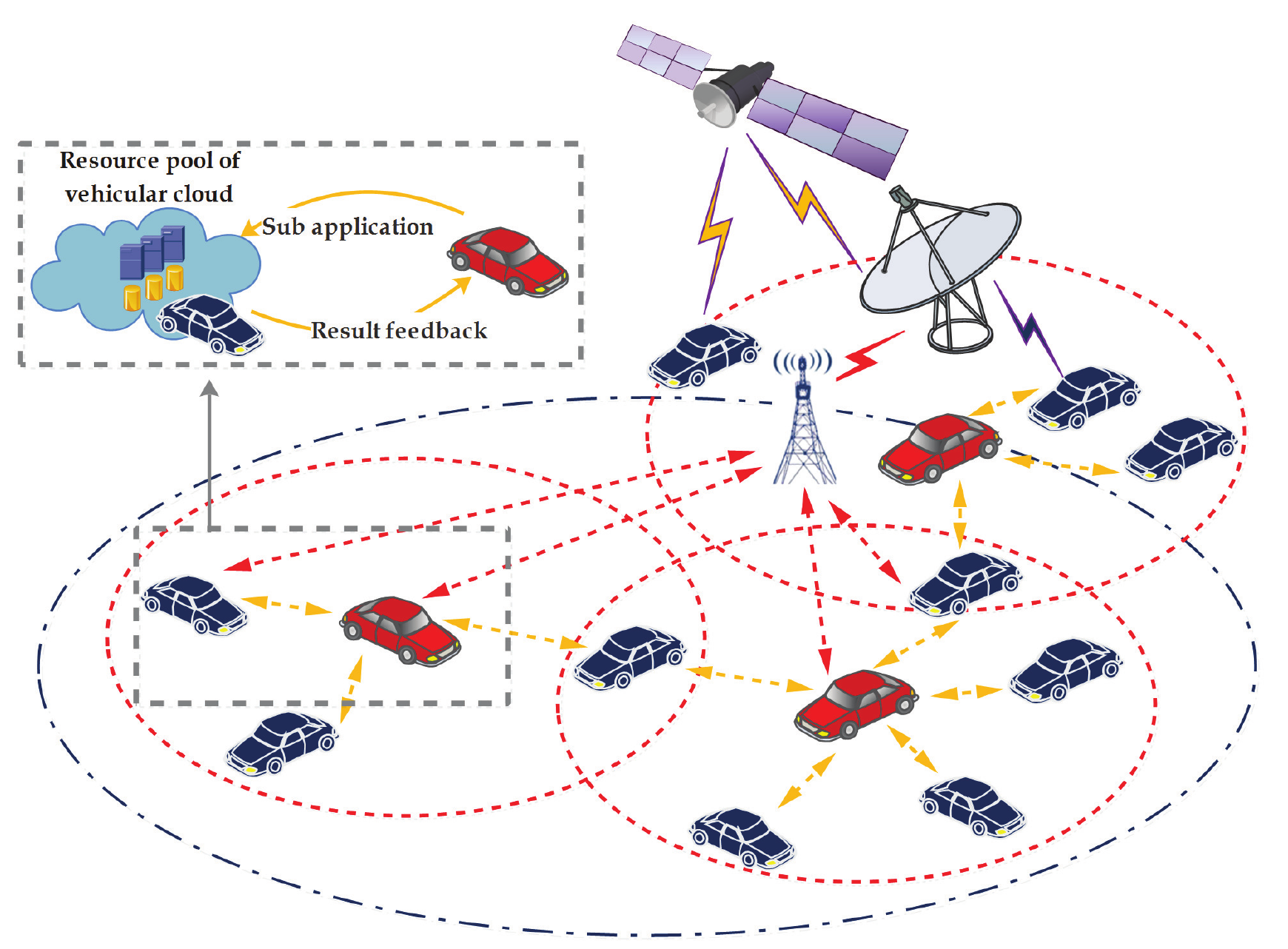}
	\caption{Schematic diagram of local interaction and computation offloading.}
	\label{fig02}
\end{figure*}

\subsection*{System Models}

The proposed terrestrial scenario consists of $B $ buyers $b_{i}\in \{b_{1},b_{2},\dots ,b_{B}\}$ and $S$ sellers $s_{j}\in \{s_{1},s_{2},\dots ,s_{S}\}$ in which buyer $b_{i} $ and seller $s_{j}$ can be denoted as tetrad $b_{i}=\{ D_{i},C_{i}^{B},T_{i}^{\max},P_{i}^{B} \}$ and $s_{j}=\{ C_{j}^{S},CC_{j}^{S}, c_{j},p_{j}^{_{Satisfy}} \}$,  respectively. $D_{i}$, $T_{i}^{\max}$ and $P_{i}^{B}$ are defined as the data size, the tolerant 
completion time of the computation-intensive application and the cost limitation of vehicle $b_{i}$ respectively; 
$C_{i}^{B}$ and $C_{j}^{S}$ are the computational capabilities (RB per second) of 
the buyer and seller; $CC_{j}^{S}$ represents the idle resources at that time,  
and $c_{j}, p_{j}^{_{Satisfy}}$ are the unit cost and satisfied unit price of seller $s_{j}$. It is worth noting that if the price paid by a buyer 
is less than $c_{j}$, then $s_{j}$ will no longer provide service for the said buyer 
due to its loss. As the offloading mechanism and pricing strategy of each group must follow the same guidance of the central controller, our discussion below is focused on only one group and the members in it. Assuming there 
are $m$ sellers in $b_{i}$'s group, the application of buyer $b_{i}$ can be denoted as $X_{i}=\sum\nolimits_{k=1}^m x_{k} +x_{B}$, where $x_{k}$ and $x_{B}$ represent the workload assigned 
to seller ${s}_{k}$ and local execution respectively. 

\textbf{Vehicular Mobility Model:} it is assumed that $B+S$ vehicles are moving in a network $\Omega_{N}=[0,\sqrt{{(B+S)}/\mu }]$, where $\mu $ is the density of vehicles per kilometer on both the east-west and 
south-north bound roads. Vehicles move according to mobility process $Q$. 
Assume that $L_{i}\left( t \right)$ and $L_{j}(t)$ 
denote the locations of vehicles $i$ and $j$, and the mobility process of a 
vehicle is stationary and ergodic such that location $L_{i}\left( t \right)$ 
has a uniform stationary distribution in the network scenario~\cite{19}. Moreover, the 
mobility processes of vehicles are i.i.d (independent and identically 
distributed). We call one contact event ${\Upsilon }_{C}$ between two 
vehicles, which occurs during $t\in [t_{1}, t_{2})$ if the following conditions are satisfied: $\| L_{i}( t_{1}^{-} )-L_{j}( t_{1}^{-} ) \|>R$, $\| L_{i}( t )-L_{j}\left( t \right) \|\leq R$ and $\| L_{i}( t_{2} )-L_{j}( t_{2} ) \|>R$, where $ R $ 
represents the transmission radius. Assuming that vehicles maintain a uniform linear 
motion during a small offloading period, the contact duration ${\Delta t}^{C} $ can be calculated easily. 

\textbf{Communication Model: }A pair of vehicles can communicate with each 
other at time $t$ when their locations satisfy $\| L_{i}\left( t \right)-L_{j}\left( t \right) \|\mathrm{\leq }R$. All 
vehicles will report information such as velocity, direction, as well as 
other parameters gathered by RSUs to the central controller periodically, which is further routed through satellite-ground stations to 
the satellite network. Due to the mobility of vehicles, different 
channel conditions lead to direct 
differences in the data transmission rates of $m$ links, represented as 
$r_{k}\in \{r_{1},\dots ,r_{m}\}$, where $r_{k}$ is the data 
transmission rate between buyer $b_{i}$ and seller ${s}_{k}$, and can be regarded as a fixed average value related to
 several factors including channel
condition, packet loss retransmission, transmission power and
the outage probability. The delivery duration of the corresponding sub-application 
content $x_{k} $ can be calculated as $t_{k}^{_{Trans}}=x_{k}/r_{k}$.

\textbf{Computation Model: }Considering a situation where an application 
is computation-intensive, and the V2V channels are unavailable, the 
local processing duration is typically smaller than or equal to $T_{i}^{\max}$, which can be denoted as ${KX}_{i}/C_{i}^{B}\leq T_{i}^{\max}$, with $ K $, a constant that serves as a maping between the data size and 
computational RBs. When the V2V channels are available, the sub-application execution duration at seller $s_k$ can be described as 
$t_{k}^{exec}=K x_{k}/C_{k}^{S}$ while 
$t_{B}=K x_{B}/C^{B}$ is the local execution time.

Overall, the sub-application duration for the seller 
${s}_{k}$ is ${t_{k}=t_{k}^{Trans}+t}_{k}^{exec}$; 
correspondingly, the total application completion time can be obtained as
$T_{i}={\max} \{t_{1}, t_{2},\dots ,t_{m},t_{B}\}$.

\section*{The joint optimization framework of incentive opportunistic offloading under delay and cost constraints}

In this section, a joint optimization problem is modeled under 
delay and cost constraints. Due to different user preferences, we 
introduce weight factors denoted as $\omega_{1}$, $\omega_{2}$, and $\omega_{3}$ to emphasize either the application completion time, monetary cost, or the incentive mechanism. For 
notational simplicity, we create four diagonal 
matrices: $\bm{H} = diag(1/{r_{1}+ K/C_{1}^{S}},\dots,1/{r_{m}+K/C_{m}^{S}}, {K/}C^{B})$; $\bm{P}^{\bm{_{Satisfy}}}=diag(p_{1}^{_{Satisfy}}, \dots, p_{m}^{_{Satisfy}},0)$; $\bm{X}=diag(x_{1},\dots ,x_{m}, x_{B})$; and $\bm{P}=diag(p_{1},\dots ,p_{m},0)$ is defined as the knockdown price per RB for each seller. Constraints are defined as follows: a) available idle resource constraint $ Kx_{k}- { CC}_{k}^{S}\leq 0$; b) contact duration constraint $t_{k}- {\Delta t}^{C}\leq 0$; c) monetary cost constraint $\sum\nolimits_{k=1}^m {p_{k}x_{k}{ - P}_{i}^{B}} \leq 0$; d) data size constraint $x_{B}+\sum\nolimits_{k=1}^m {x_{k}=D_{i}} $; e) non-negative constraint ${\forall k\in \left\{ 1,2,\dots ,m \right\},x}_{k}\geq 0$ and $p_{k}\geq 0$; and f)~incentive constraint ${CC}_{k}^{S}= 
0 \,\,{\rm if~} { p}_{k}\leq c_{k}$. %\\ 
%{CC}_{k}^{S} &  {\rm otherwise} \\ 
%\end{cases} $. 
%\begin{cases}
Thus, the objective 
function for each cooperative group that includes one buyer and $m$ sellers can 
be specified by using infinite-norm, 1-norm, and F-norm as in the following 
optimization problem: 
\begin{gather}\label{key}\nonumber
\Phi = \arg\min_{X,P}{\omega_{1}\left\| \bm{HX} \right\|_{\infty }}{ +\; \omega_{2}\left\| 	\bm{PX}\right\|}_{1} + \omega_{3}\left\| \bm{P}-\bm{P^{Satisfy}} \right\|_{F}\\
{\rm s.t.~ a),~ b),~ c),~ d),~ e),~ f)}
\end{gather}
The first part of the objective function has the physical significance 
of reducing application completion time. The monetary cost constraint is shown as  
the second part, and the incentive mechanism is reflected in the third part which 
means the closer the knockdown price is to the seller's satisfied 
price, the more willing a seller will be to provide service in the 
future. 

\textbf{Solution: }Owing to the fact that the objective function (1) with constraints is 
non-convex and NP-hard, a simulated annealing algorithm is utilized to obtain feasible solutions. Simulated annealing is a heuristic algorithm based on the Markov Chain Monte Carlo (MCMC) as well as the metropolis 
criterion, which can lead to high-quality and cost-effective approximations 
of global optimal solutions. 

\section*{Performance evaluation}

\begin{figure}[h]
	\centering
	\includegraphics[width=\linewidth]{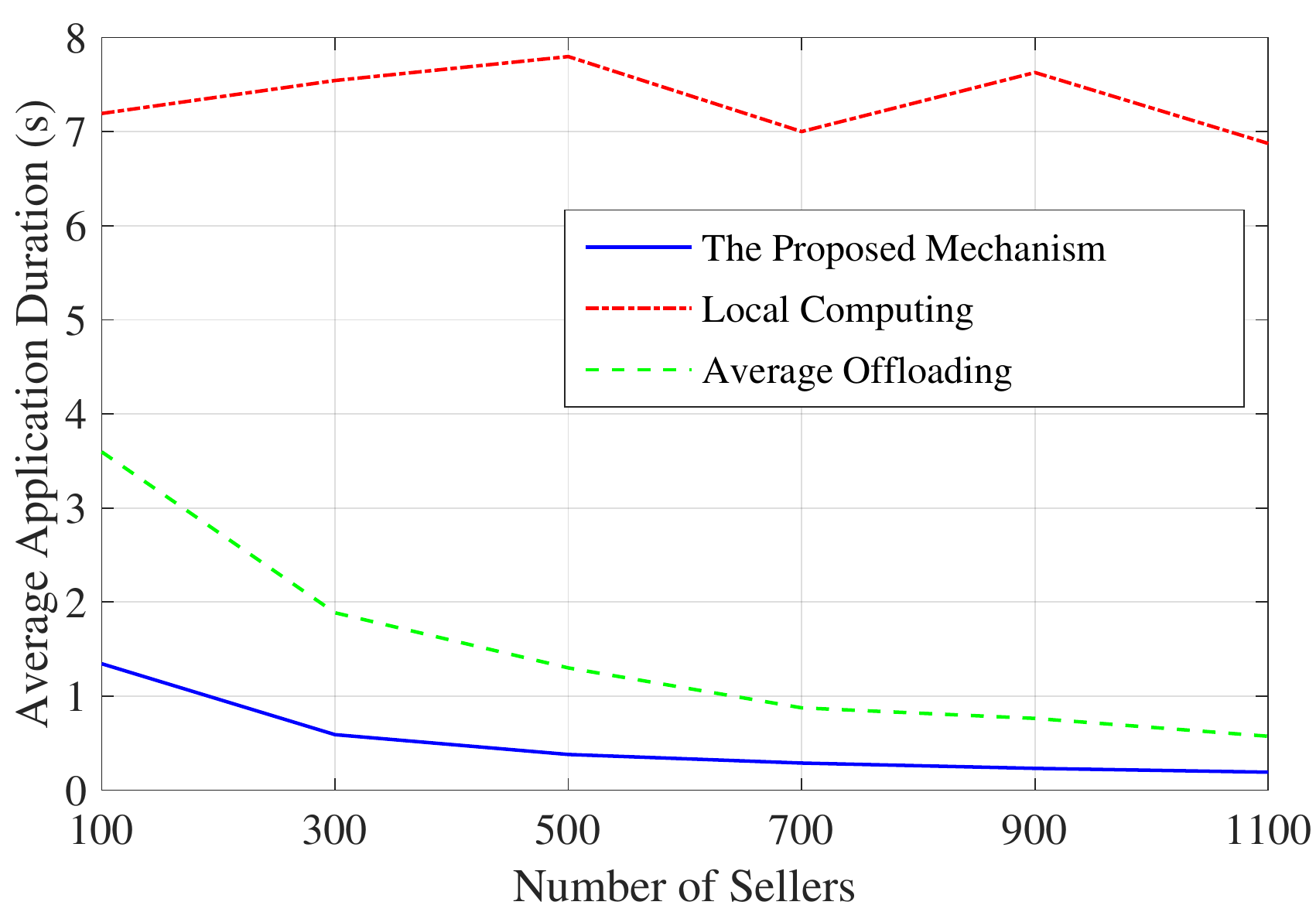}
	\caption{Average application completion duration with the number of sellers.}
	\label{fig03}
\end{figure}

Simulation results are presented in this section containing 100 buyers and multiple sellers (from 100 to 1100) in a wide coverage area under the satellite-ground 5G IoV framework. The communication range of a vehicle is set to be $R$ = 250m; the application data size D$\in$[5Mb, 6Mb] and data transmission rate r$\in$[3Mb/s, 6Mb/s]. The weight parameters are set to be $\omega_1 =\omega_2 = \omega_3 =1/3$.  Figure 3 shows the average application completion duration with the number of sellers by comparing the proposed mechanism with two other schemes: the Local Computing algorithm where one application is totally executed locally without offloading, and the Average Offloading algorithm, where one application is distributed evenly to every seller in the cooperative group under contact duration constraints. As can be seen, the Local Computing algorithm has the highest application completion duration, whereas that of Average Offloading decreases as the number of sellers increases but still remains higher than that of the proposed mechanism without considering different computing capacities among sellers. The proposed mechanism demonstrates the best performance by jointly accounting for computational capacities, contact duration, and different channel conditions among the sellers.

\begin{figure}[h]
	\centering
	\includegraphics[width=\linewidth]{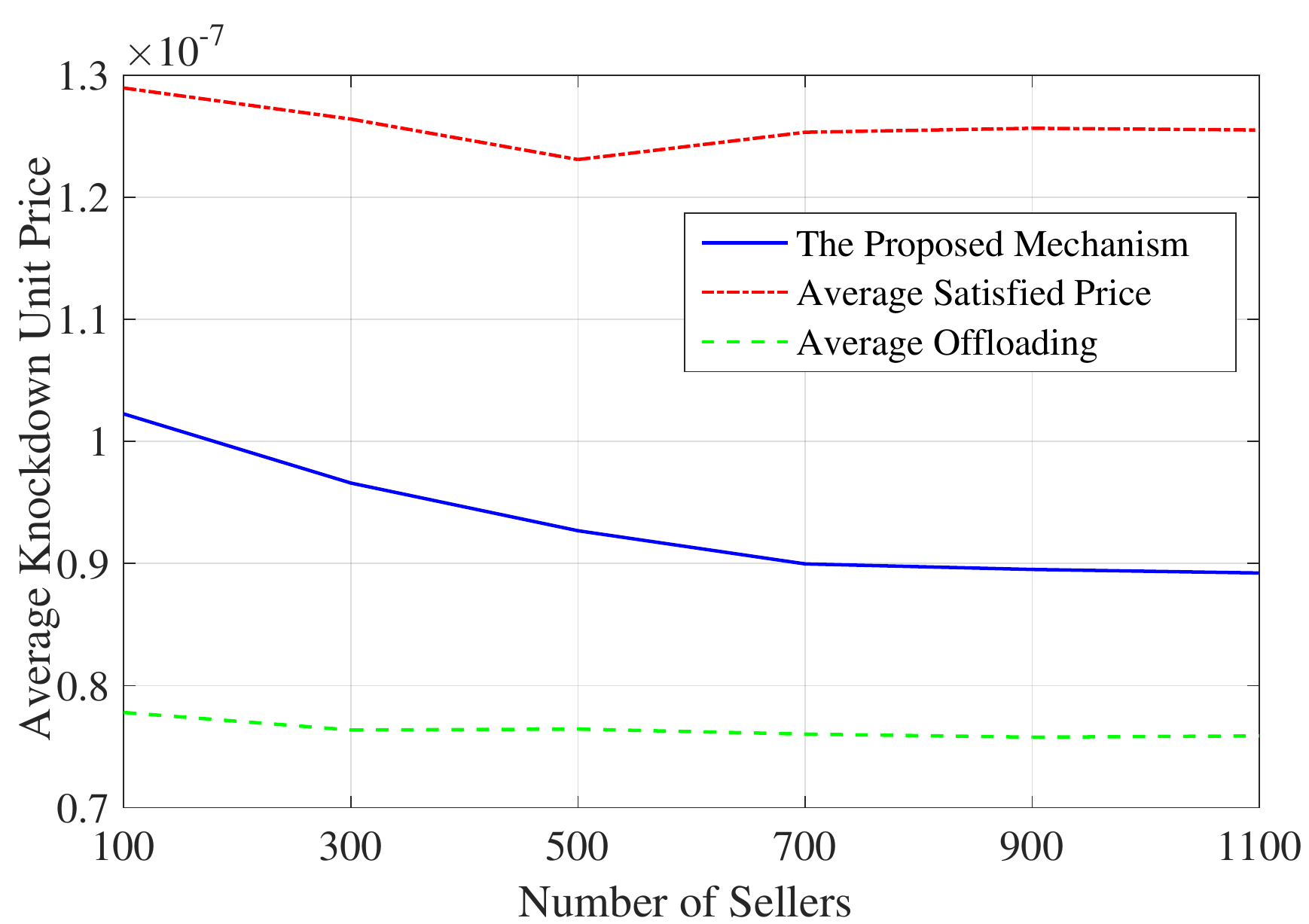}
	\caption{Average knockdown unit price with the number of sellers.}
	\label{fig04}
\end{figure}

Figure 4 shows the average knockdown unit price with the number of sellers. Due to the competitions among the sellers, the knockdown unit price trends downward in the proposed mechanism but still is much higher than that of Average Offloading, where the knockdown price is only slightly higher than the seller's cost without an incentive mechanism. In other words, central controllers in the Average Offloading scenario are not concerned with how much sellers earn as long as they do not experience a loss. In contrast, sellers in the proposed framework receive better payments and are more willing to provide idle resources for sustainable and green development of the satellite-ground IoV.

\begin{figure}[h]
	\centering
	\includegraphics[width=\linewidth]{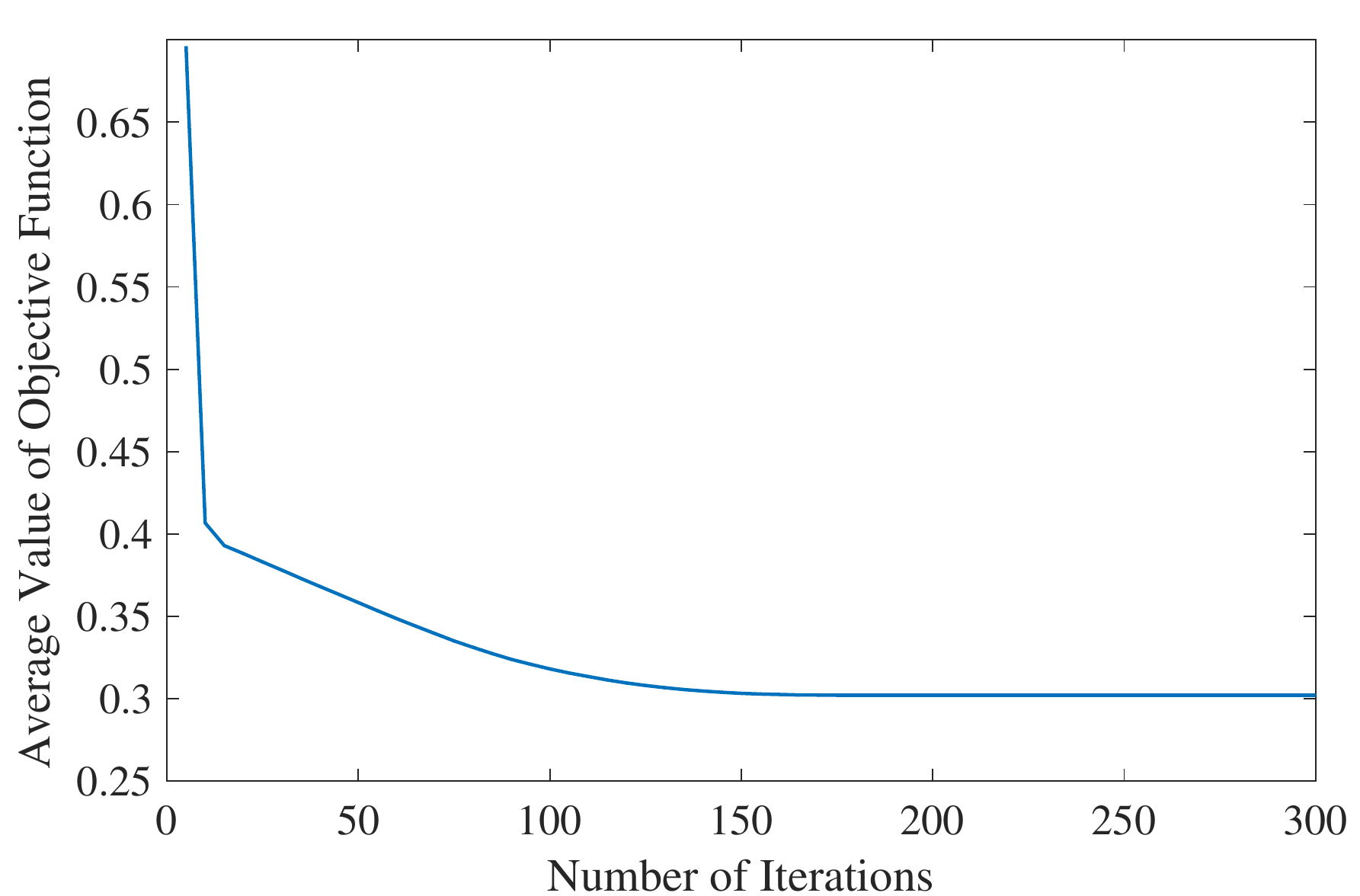}
	\caption{Convergence of the simulated annealing algorithm.}
	\label{fig05}
\end{figure}

An example of convergence in the simulated annealing algorithm is shown in Fig. 5, where 100 buyers and 600 sellers are tested for the rate of convergence, which is regarded as a critical factor in a rapidly changing environment like IoV, especially when vehicles communicate with each other through opportunistic V2V channels. As illustrated in Fig. 5, the algorithm begins to converge when the number of iterations exceeds 100, which corresponds to a negligible period with the current computing technology; therefore, the offloading decision-making at the central controller can handle the high mobility of vehicles effectively.
\section*{Conclusion}

In this article, we devise an integrated architecture of satellite 
networks and 5G Internet of Vehicles that effectively provides both seamless 
coverage and resource management from a macroscopic point of view. Then, an incentive mechanism based 
joint optimization problem for computation offloading among vehicles is 
modeled under delay and cost constraints, where a service buyer can 
significantly reduce the application completion duration and control monetary costs while 
service sellers are motivated to promote
sustainable green network development. Using a simulated annealing 
algorithm, simulation results are presented to substantiate the practical 
effectiveness and efficiency of the proposed mechanism.

\ifCLASSOPTIONcaptionsoff
  \newpage
\fi

\vfill


\end{document}